# Can extremism guarantee pluralism?

**Floriana Gargiulo**[1,2], **Alberto Mazzoni**[1]

1- Institute for Scientific Interchange (ISI) Foundation, Turin, Italy
2- Complex Network Lagrange Laboratory (CNLL)

## Abstract

Many models have been proposed to explain the opinion formation in a group of individuals; most of these models study the opinion propagation as the interaction between nodes/agents in a social network.
The opinion formation is a very complex process and a realistic model should also take into account the important feedbacks that the opinions of the agents have on the structure of the social networks and on the characteristics of the opinion dynamics.
In this paper we will show that associating to different agents different kind of interconnections and different interacting behaviour can lead to interesting scenarios, like the coexistence of several opinion clusters, namely pluralism.
In our model agents have opinions uniformly and continuously distributed between two extremes. The social network is formed through a social aggregation mechanism including the segregation process of the extremists that results in many real communities. We show how this process affects the opinion dynamics in the whole society.
In the opinion evolution we consider the different predisposition of single individuals to interact and to exchange opinion with each other; we associate to each individual a different tolerance threshold, depending on its own opinion: extremists are less willing to interact with individuals with strongly different opinions and to change significantly their ideas.
A general result is obtained: when there is no interaction restriction, the opinion always converges to uniformity, but the same is happening whenever a strong segregation process of the extremists occurs. Only when extremists are forming clusters but these clusters keep interacting with the rest of the society, the survival of a wide opinion range is guaranteed.

# 1- Introduction

There are a hierarchical and a horizontal way of opinion interaction. Opinion interaction is hierarchical when a single powerful agent is modifying the opinion of large sectors of the community – religion, media, governments often play this role. We are not going to consider this kind of interaction in our present work. Horizontal interaction occurs when the dialogue between two or more members of the community results in a change of the opinions of the debaters. This kind of opinion dynamics is the one we are going to study in the present work and is based on the principle of social influence: the more two persons interact, the more similar they become.
Horizontal interaction has been considered less important in a society in which the opinion is spread by means of powerful centralized media. In recent years, however, a u-turn in the media structure has been observed: with the advent of the internet era, grassroots autonomous media structures emerged [1,2]; blogs, discussion lists, web communities are created, giving to the social influence principle a novel important role in the opinion evolution mechanism.

Many mathematical models of the social influence principle have been proposed to describe how two persons make their opinions more similar after a discussion (for a review see [3]). Some models, like for example the Voter model [4,5], describe the exclusive choice between two possibilities, like the case of a referendum or of a single winner election. More refined descriptions like the Axelrod's model [6] define the opinion as the resultant of a set of cultural traits. Finally the Deffuant model [7] describes the more complex setup of continuously differentiated opinions. In our model we adopt a particular form of this opinion distribution: there are two opposite positions and a continuous range of intermediate opinions. This could be for instance the distribution of the opinions regarding a single controversy (the decision for a city to stop traffic on Sundays, or for a country to enter a war, for instance). We are going to call for the sake of simplicity "extremists" the people having an opinion close to one of the two extremes and "neutrals" those who do not take side. We would like to stress that "neutral" does not mean "moderate", for instance belonging to a centre party in the left-right arrangement of political parties in Europe. "Moderates" have usually very strong opinions that they are willing to defend and spread. In the context of our model we consider "neutrals" people that do not have opinions, at least on selected topics. From the point of view of elections, neutrals are those who do not vote because they do not care about, not those who are voting for the moderate parties.

Deffuant [7] introduces also the concept of bounded confidence: two persons interact only if the distance between their opinions is lower than a given tolerance threshold. Furthermore, the opinions converging mechanism due to the social influence, in the model, is a symmetric drift of both the agents in the direction of an intermediate opinion.

In this paper we would like to introduce a more complex but to our advice also more realistic description. In our model the tolerance threshold and the drift indicator are not considered uniform values over the entire population but are rather different for each person [8] according to its opinion [9]: the stronger the opinion of an individual, the more it will prefer to discuss only with people sharing similar ideas. A neutral position on a certain topic is usually determined by a lack of knowledge that would lead neutrals to interact with a wider range of opinions and to be easily persuaded by stronger viewpoints. On the other side, extremists are less tolerant and much more confident of their opinions (like the inflexible minorities in [10]).

Deffuant model has been applied to several kinds of social structures: from all-to-all networks in which every couple of agents can interact with each other, to lattices where the number of neighbors with whom is possible to interact is fixed [7], to networks where the number of neighbors has a more complex distribution [11].
In all these cases the structure of the network was considered only as the framework for the opinion evolution and it was not influenced by the opinion of the single social actors: the extremist could a priori have the same connections (both qualitatively and quantitatively) as a completely neutral person. This situation is not realistic since in many different fields, and in many different social networks a strong segregation effect is observed: in urban areas such as in web communities, people with very similar traits that are too distant from the average of the community, tend to build their own sub-communities. [12,13].

In our model we take into account also the effect of the opinion on the network structure, introducing a social network construction mechanism describing segregation process. The mechanism that leads individuals to choose acquaintances with very similar cultural, religious or racial traits, causing the formation of strongly connected opinion communities is called homophily [14,15,16 ,17]. We will model this phenomenon introducing an opinion dependent homophily: during the first phase of network formation extremists will choose to be more likely surrounded by persons sustaining their opinions while neutrals will not have prejudice in forming links with anybody. The architecture of the links is fixed and is not changing during the opinion evolution, occurring when the network is completely formed. The architecture is then determined by the different starting opinions. This situation correspond to a rapid opinion evolution during which the social links are not rewired.

The paper is structured as follows: in section 2 and 3 we describe the algorithms that we use to construct the interaction network and to model the opinion dynamics.
In section 4 we show the numerical results of the opinion propagation: how the opinion dependent homophily in the network structure and the opinion dependent tolerance in the opinion dynamics can lead to different convergence scenarios. We observe that in societies where the extremists self-segregate pluralism is never achieved, because when extremists are less confident in their ideas, the opinion of all the agents always converge to an uniform opinion, while if they are more confident, only very small and isolated groups of extremists survive. Pluralism is reached only when the extremists are integrated in the society but are simultaneously strongly convinced of their ideas: this is the only case in which many different opinions persist at the end of the simulation.

## 2- Methods 1: Social structure

Each simulation is divided into two parts. First, the social structure network is formed, i.e. we determine the links connecting the different individuals. Then, the structure is keep fixed and we let the opinions of the individuals interact, but no link is formed or destroyed. In this section we describe the formation of the social network.
If the number of agents was sufficiently small, i.e. if we were dealing with a small social aggregation (the students of a class, the colleagues of a small company, a small village…) it would have been a good approximation to consider that all the agents interact with each other. In this case the agents form a complete clique: the structure of a similar society is a completely connected graph and all the nodes have the same number of connections (degree). The fact that all agents interact does not mean that all interaction are identical: different intensities can be associated to different links but in this paper we are not going to deepen on this topic.
We are not adopting this approach, since all-to-all connections models can not be used for bigger societies where the agents are not able to interact directly with all the others. However we will show in section 4.2 that our results are valid also in this kind of networks.

In many realistic social networks a strong heterogeneity in the degree is observed: most of the agents have a small number of contacts but there is also a significant number of nodes with many connections.

This behaviour can be explained, for example, with the preferential attachment model introduced by Barabasi and Albert [18]: starting from an initial core, at each step a new agent gets connected to *m* pre-existing agents. The targets of the new connections are selected with a probability that is proportional to the degree of the old nodes. In this way, along the growth process, the most connected agents will become more and more connected, while the more isolated ones remain always less attractive.

The idea is that persons with more relations are more likely to have new relations: your friends introduce to you new friends, while when you are in an environment with no friends it is difficult to start. Such a procedure generates a degree distribution that, when the number of nodes becomes big enough, is a power-law with exponent $\gamma=-3$: $P(k)=k^{-3}$ (where k is the degree, and P(k) the probability of finding nodes with degree k).

The Barabasi-Albert model provides a basic description for many natural and artificial systems but it is not fully exhaustive for modelling social interactions.

The process we would like to describe in this paper is the use of social attitude, and in particular opinion affinity, as a preferential mechanism for building social connections [19]: in social frameworks individuals who share common interests and ideals tend to be strongly connected between them and to form distinct communities. In particular, a phenomenon that is largely observed in various social structures and at all the possible scales (from internet community to big cities structure) is the segregation process: famous example at urban level are the "Ghettos" in the US metropolis and the "Banlieues" in the French cities that are communities built on social and economical stratification.

Segregation is a mechanism that tend to favor the creation of links between those members of a society whose characteristic traits are distant from the average of the community.

Three different mechanisms can generate segregation. Segregation can be imposed by external factors: in many countries there are laws allowing only some specific ethnic communities to live in given areas, or confining other communities, and rent price gradient has often practically the same effect. A mixed network can spontaneously become a become a segregated one thanks to rewiring [20], for example when people choose to live among those who share similar opinion ("The Big Sort" in The economist, June 19[th] 2008).

Finally, segregation can also result directly by the same generation process of the social network, in this case we can speak of self-segregation like, for example, in the case of ethnic or religious communities or some extreme political organizations.

We consider this last mechanism and we build a network model according to these assumptions. Self-segregation is generated by the fact that individuals that are far from the main opinion in their community (*extremists*), rely much on their identity in social interactions, so choose a criterion strongly based on the homophily to select their contacts. On the other side agents with neutral opinion, do not have a priori prejudices in choosing new connections. In order to model the segregated structure, we keep into account the dependence of the homophily on opinion and, in the inclusion mechanism of a new agent, we consider a probability function that depends on these different attitudes.

We consider a continuous opinion model where the opinion of each agent is randomly extracted in the range [-1,1]. We start form an initial core of randomly connected nodes and then, at each step of the network formation process we connect a new node to *m* pre-existing nodes with the following algorithm:
- A random opinion is associated to the new agent, $o_N$

- The new agent connects to the pre-existing agents with a probability:

$$P_{(N \to i)} \approx k_i * Exp[-\beta |o_N||o_N - o_i|] \qquad (1)$$

where $k_i$ is the degree of the pre-existing agent.

The exponential function that appears in the connection probability (that is displayed in figure 1) can assume values in between 1 and 0 and reproduces the requests that we listed before. The "homophily parameter" $\beta$ modules how agents are strict in the choice of their relational approaches: if $\beta=0$ the model reduces to the Barabasi-Albert network where the extremists are completely inserted inside the society. For $\beta>0$ the probability of interacting between two agents is maximal if they have exactly the same opinion (since $o_N-o_i=0$ and the exponential becomes 1) and decreases (with a slope depending on $\beta$) for more distant opinions. For large values of $\beta$, the extremist will connect only to the very similar agents, creating their own communities and avoiding the mixing with the rest of the society. On the opposite, independently from $\beta$, the neutral agents, will not have any preclusion to link to anybody because when $o_N=0$ the exponential is equal to the maximum value 1 for any value of the other parameters.

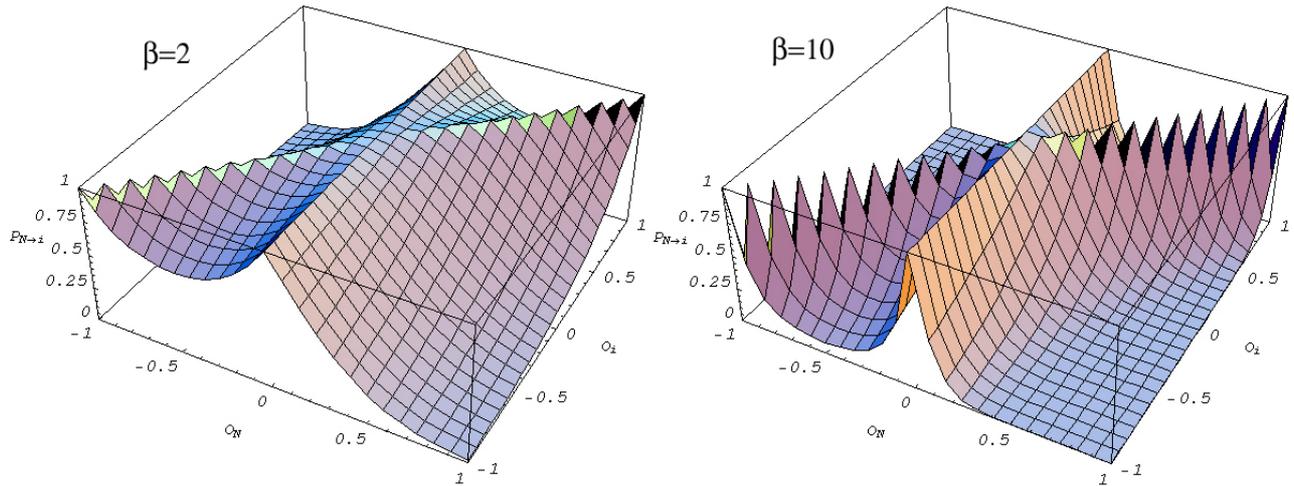

**Figure 1:** Homophily in network formation. Probability of connection between the new node (*N*) and a node (*i*) already included in the network as a function of their opinion. The left plot is realized with a parameter $\beta=2$ in Equation 1, and the right one for $\beta=10$.

As reported in figure 2, such kind of construction gives rise to a scale free network, i.e. to a power law distribution of the degree probability, with a cut-off appearing only for extremely high values of $\beta$ ($\beta>20$). But even if such process preserves the same degree distribution of the Barabasi-Albert network, the clustering structure changes with increasing values of the parameter. To measure the level of cliquishness in the graph we used the following clustering coefficient: for each vertex the clustering coefficient is given by the effective number of links between the neighbourhood divided by the total number of links that could possibly exist between all of them. This measure is one if the neighbours form a complete clique, while it is zero if they are completely disconnected.
As can be noticed in the figure the clustering coefficient increases with $\beta$, suggesting the creation of more and more separated opinion communities.

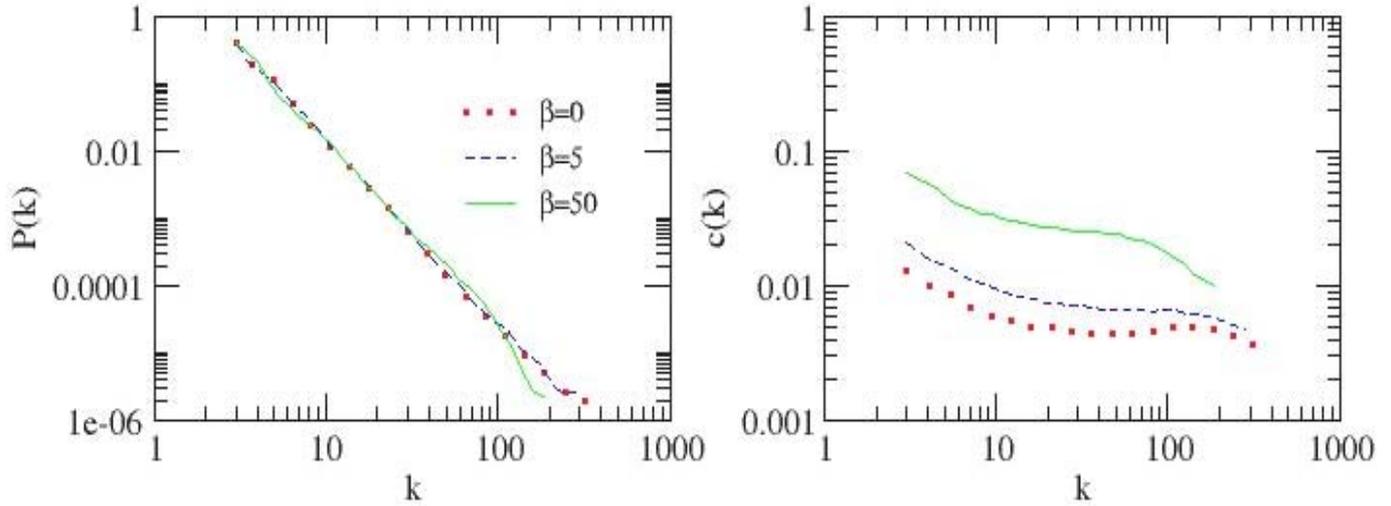

**Figure 2:** Network structure. Degree distribution (left plot) and clustering coefficient as a function of the degree (right plot) for a network of 10000 nodes and for different values of β. The result is averaged over 100 realizations of the network.

The discussion on community detection inside a network is a popular topic in the recent network theory literature [21,22]. We are not interested to deal with such topological structure, so we use a very simple tool to analyze if the network effectively shows a segregated structure according to the opinion dependent homophyliac process that we described before. In figure 3 we plot the opinion couples ($o_i$,$o_j$) for all the possible i,j neightbours (black points). In the same graph we plot the binned data for the opinion versus the average value of the opinion of neighbours, for different values of β (continuous line).
As we can notice, for β>0 the extremists agents neighbors, on average, have their same opinions; in this sense we say that the extremists tend to self-segregate.

The process we have described sets the initial conditions for the opinion evolution. It is important to stress that in the present network the social network structure is stable, i.e. links remain the same thorough the opinion evolution, even if their strength and efficiency can vary, as described in the next section.

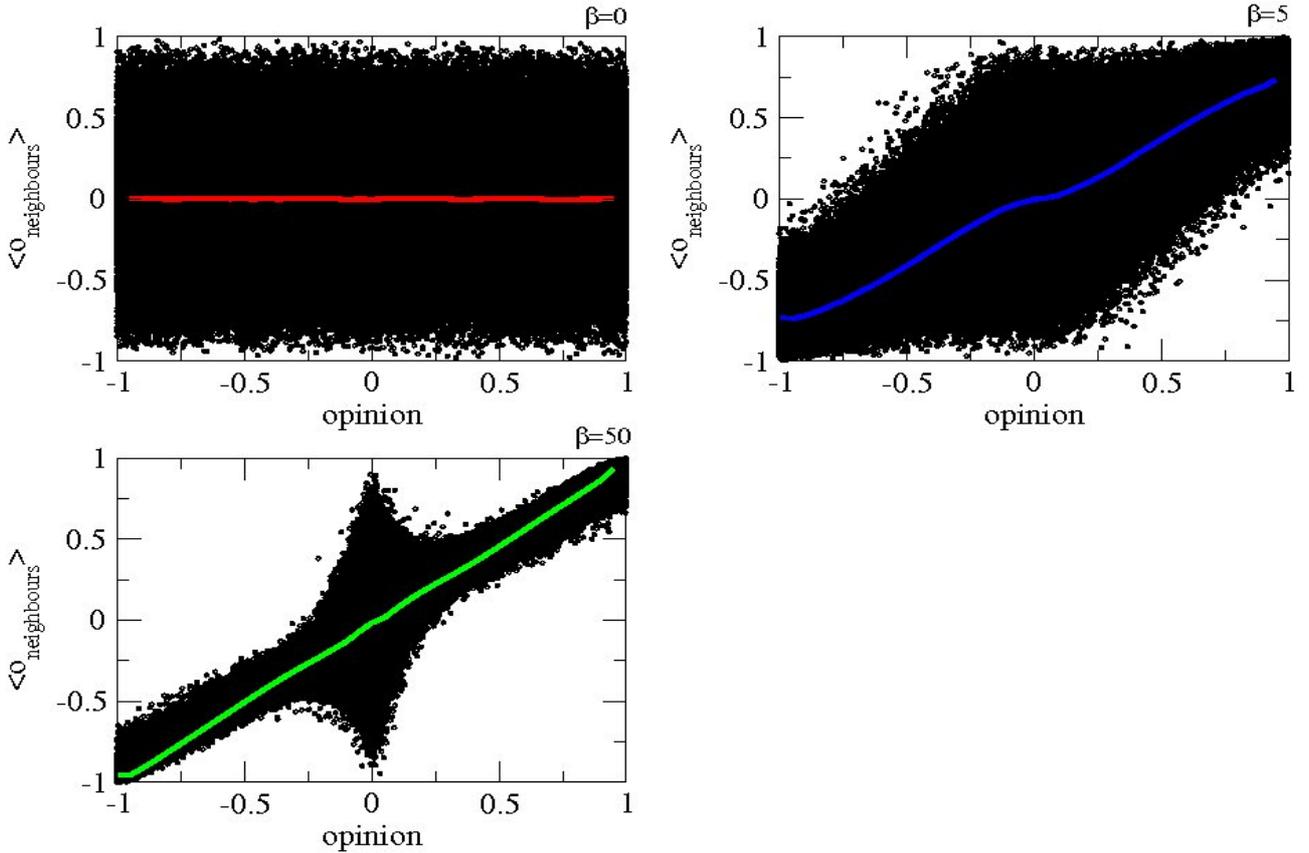

**Figure 3:** Neighbours opinion at the end of network formation. The continuous line represents the average opinion of the neighbours of an agent as a function of the opinion of the agent. The three plots correspond to three different values of the parameter. The black points represent all the couples ($o_i$, $o_j$) for all the links of the network.

## 3- Methods 2: Opinion Dynamics

Once that the network is formed, we let the individuals interact, i.e. at each step the opinion of each individual changes, becoming more similar to the opinion of its neighbours with some conditions that are the subject of this section. When all the opinions have been updated the interactions start again until a stable opinion configurations is obtained. This is a well known procedure [7].
The modification that we introduced in our model aimed to take in account that not all the individuals interact have the same efficacy in changing the other individuals opinion, and this is often due to the strength of the starting opinions of both parties.
Nixon introduced the idea of "Silent majority" (3 nov 1969) to convince US citizens that all of those who were not expressing an opinion on Vietnam war were in fact supporting the war. In a correct picture, however, if an individual is neither expressing nor acting to support any opinion on a particular topic, we must consider this individual with no opinion, neutral. This is usually caused by a mix of a lack of interest, due to the perception that involvement is not going to cause any consequence on one's life, and a lack of information on the different positions.
We try to explain the ideas originating our model of opinion propagation with the example of a presidential campaign. When neutrals, people who have not yet taken position, get in touch with strongly motivated people, supporters of one candidate, they can easily change their mind since they do not have any firm idea. But i) the motivated people are not going to weaken their opinion after this interaction, so the interaction is asymmetric, ii) if it is relatively easy to take a position, dismissing it once that one is convinced is far harder, so sensitivity depends on the strength of one's opinion. Finally, in the presidential campaign example, it is clear that if neutrals will initially listen to everybody to decide with position to take, and everybody is going to try to convince them, supporters of a party will not waste time going to discuss in other parties headquarters, so iii) the range of interactions is larger when the opinion is weaker and vice versa.

All these assumptions on the different interacting attitudes for radicals and neutrals can be connected to a well known concept in social psychology: the social identity [23,24]. People that more strongly support a position are usually part of groups, parties or, in any case they feel to be part of a collective organism. This fact increase the self-esteem of the individual and then the strength with whom he/she defends the supported position. Simultaneously the more the identification process is strong, stronger will also be the intergroup discrimination and the denial of other ideologies.

To model these 3 factors we introduced a parameter α, ranging from 0 to 1, measuring the role of sensitivity in the interactions.

First of all we define an opinion dependent tolerance threshold:

$$t_i = 1 - \alpha |o_i| \qquad (2)$$

In this way the interaction range varies with the opinion (with a strength depending on α) to represent the interaction limitations of extremists. This is in our opinion the simplest function capturing the features we are interested in: the threshold varies linearly with both opinion and strength, perfect extremists ($|o_i|=1$) interact with nobody and perfect neutrals ($|o_i|=0$) interact with everybody.

Two agents interact only if they are connected by a link and if:

$$|o_j - o_i| < \min(t_i, t_j) \qquad (3)$$

If α =0 we have a uniform interaction range of 1 (half of the complete opinion range). In any case the range does not depend on α for neutral agents (o=0), while for instance strong extremists of both sides (|o|=1) will talk to nobody if α =1 and to a range of 0.5 if α =0.5.

The amount of the opinion modification following an interaction depends on the individual tolerance, to represent the fact that extremists are less likely to change their mind

$$\Delta o_i = t_i * (o_j - o_i)/2 \qquad (4)$$

If α =0 any interaction will end with the two nodes sharing the same average opinion, while for α>0 the modification is reduced. For any value of α, after an interaction, a neutral agent (o=0) will move half way toward the opinion of the other interacting node, but the modification of any node with |o|>0 would be progressively reduced by the (1- α|o|) factor when α or |o| is increased.

Notice that, for α =0, our model reduces to the standard Deffuant model with confidence parameter ε=1 [7].

## 4- Numerical results

The model we introduced has two main parameters: a structural parameter β that modules how the (opinion dependent) homophily influences the network construction and a dynamical parameter α that connects the tolerance of each agent to its opinion and determine the possibility to modify its opinion. We observed the effect of each parameter on the dynamics of the opinion performing simulations with different values of α and β on a population of $N_{ag}$=1000 agents. For the evolutionary process, at each time step we update, one by one and in a random order, the opinions of the single agents (asynchronous update). We will first present some qualitative results, to give an idea of the dynamics, presenting statistics in the next section.

We start considering two extremes values of α, α=0 (classic Deffuant model) and α=1 (in which opinion and tolerance are most related) for different values of β.
The results are displayed in Figures 4 and 5.

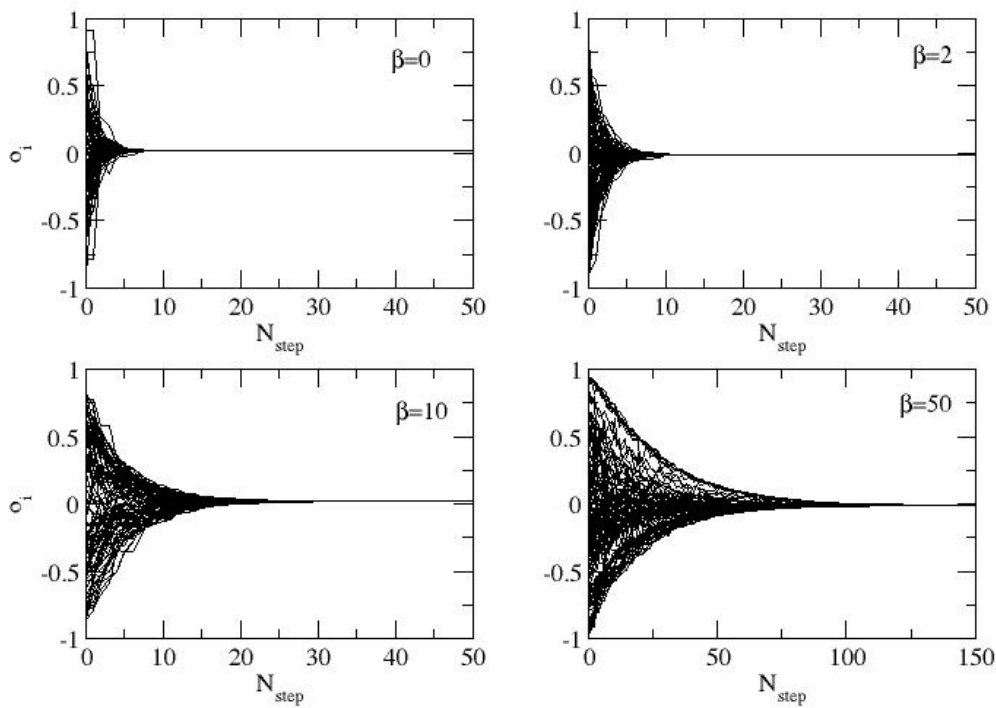

**Figure 4:** Results of opinion dynamics with α=0 (tolerance and interaction do not depend on opinion) on network structures built with different values of β. Each line represents the opinion evolution of one agent. Only a subset of 100 agents out of 1000 is displayed. The plot is the sketch of one realization of the simulation.

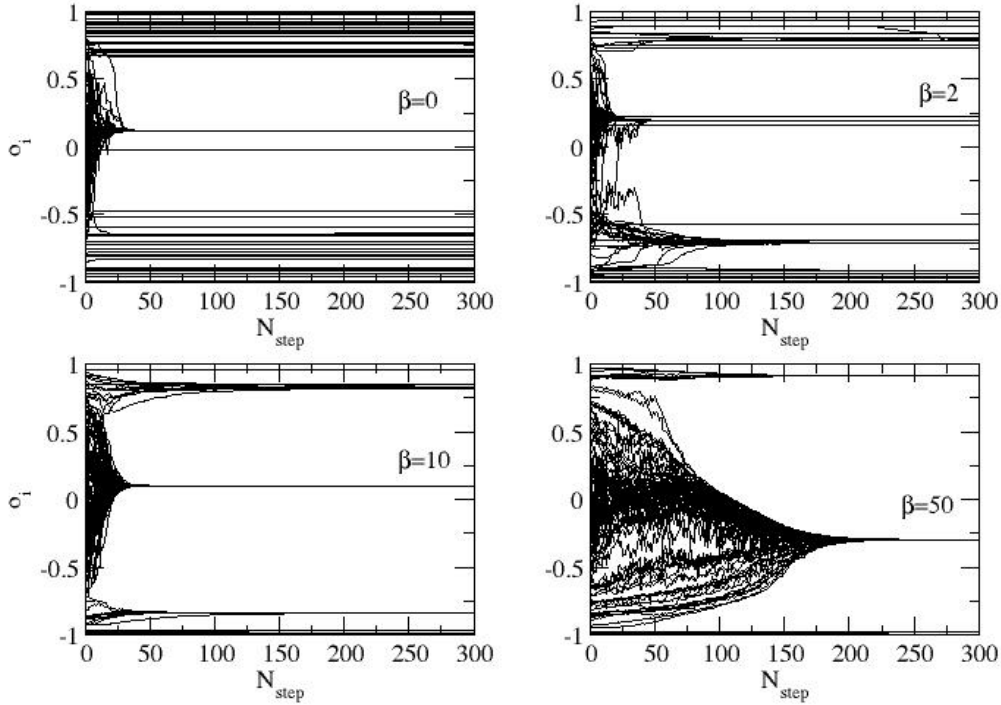

**Figure 5:** Results of opinion dynamics with α=1 (tolerance and interaction strongly depend on opinion) on network structures built with different values of β. Each line represents the opinion evolution of one agent. Only a subset of 100 agents out of 1000 is displayed. The plot is the sketch of one realization of the simulation.

For α=0 (figure 4) the opinion evolution process is exactly the basic Deffuant model with the tolerance parameter fixed to 0.5: independently from the structure of the social interactions we always obtain the convergence to the average opinion. The level of segregation of the community does not alter the final result but only the convergence time to such situation. If β is bigger, the opinion converges in a longer time because the average opinion difference between neighbours is smaller, and as consequence the average modification per step is smaller, so a higher number of steps is required to reach uniformity.

The result is completely different for α=1 (figure 5). In this case the tolerance of each agent depends on its opinion: the extremists are more convinced of their own ideas, they interact only with similar people and they feel less the bias of the other's opinion.
Under these conditions the global consensus is never reached but different behaviors can be observed depending on the segregation level. When the value of β is large (β=50), i.e. extremists are also topologically segregated, the consensus is almost reached apart from some small groups of extremists that maintain their opinion: a large central majority cluster is observed but also a small group whose opinion lies at the extremes of the interval form opinion clusters.
When β goes to zero the network structure is a Barabasi-Albert network and consequently the extremists are completely integrated with the society. In this case the consensus is far from being realized: many smaller opinion minorities are formed and a wide range of opinions is present at the end of the evolution.

Similar considerations can be done on the effect of the dynamical parameter. In Figures 6 and 7 we show the plots regarding the opinion evolution for different values of $\alpha$ for two extreme values of $\beta$, $\beta=0$ and $\beta=50$.

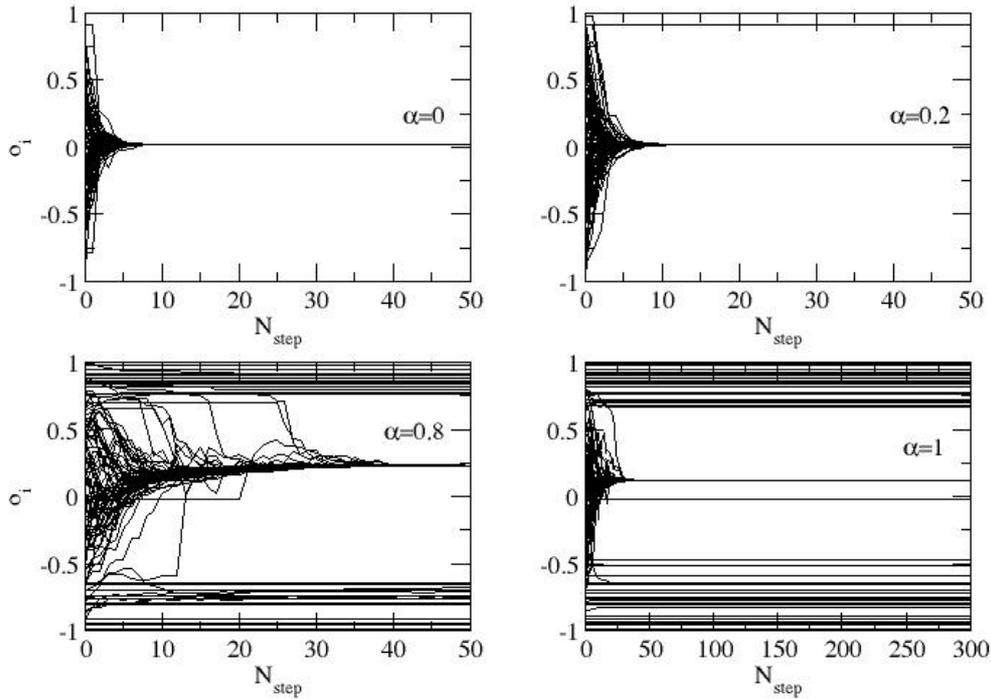

**Figure 6:** Results of opinion evolution on networks built with $\beta=0$ (Barabasi Albert network) for dynamics using different values of $\alpha$. Each line represents the opinion evolution of one agent. Only a subset of 100 agents out of 1000 is displayed. The plot is the sketch of one realization of the simulation.

If we consider an opinion independent link structure, $\beta=0$ (figure 6), we notice that, varying the parameter $\alpha$ different results are obtained: as we already observed, if $\alpha=0$ the consensus is rapidly reached. Increasing the parameter $\alpha$ leads to a larger number of opinion minorities.

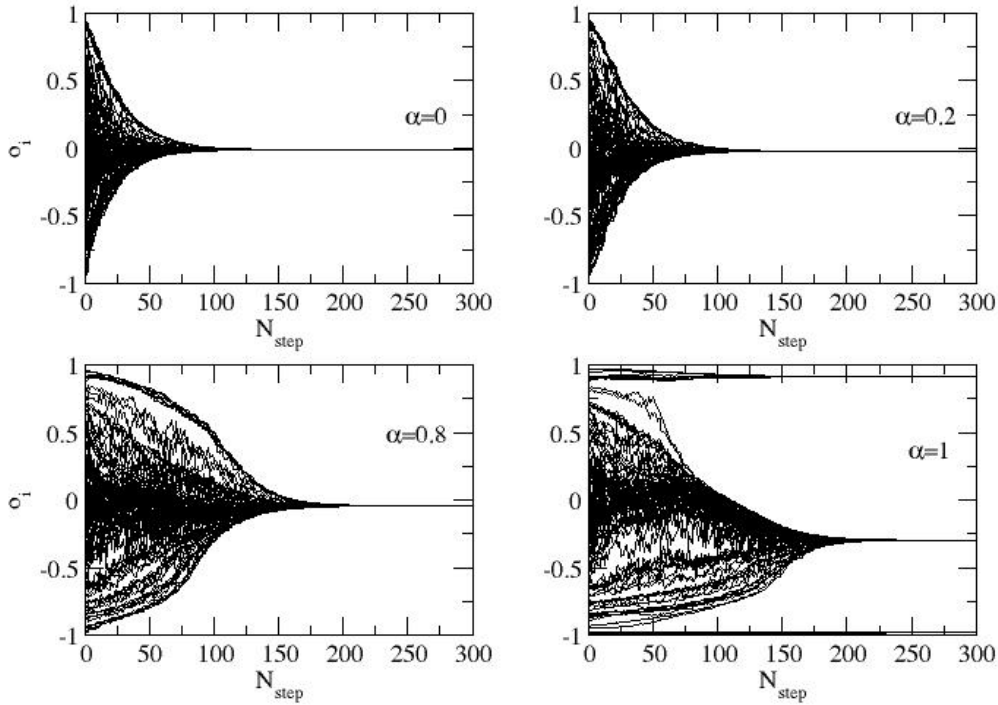

**Figure 7:** Results of opinion evolution on networks built with β=50 (strong homophily) for dynamics using different values of α. Each line represents the opinion evolution of one agent. Only a subset of 100 agents out of 1000 is displayed. The plot is the sketch of one realization of the simulation.

On the contrary, for β=50, for most of the range of α the consensus is reached and only for α=1, small groups of extremists survive the uniform consensus process.
In fact, for large values of β the network structure is strongly correlated to the initial opinions of the agents. Such correlation allows the existence of gradual paths of communication that always lead to convergence.
The survival of side clusters for α=1 (where the tolerance is strongly dependent on the opinion) is due to the fact that, in this situation, the most extremist agents are not involved in the global opinion dynamics process since their tolerance is too small to interact with someone that is far from their ideas; they just interact with the most similar agents creating a sort of opinion niches.

For large values of α a non-consensus situation is observed independently from β, but the correlation between the opinions and the network structure at the end of the simulation strongly depends on β: in Figure 8 we plot the opinion of each agent vs the opinion of her neighbours for β=0 and β=10. For β=0 the extremist agents have neighbours with all the possible available positions, while for β=10, also at the end of the simulation a segregated structure for the radicals positions is observed: they only have link between agents with very similar opinions.

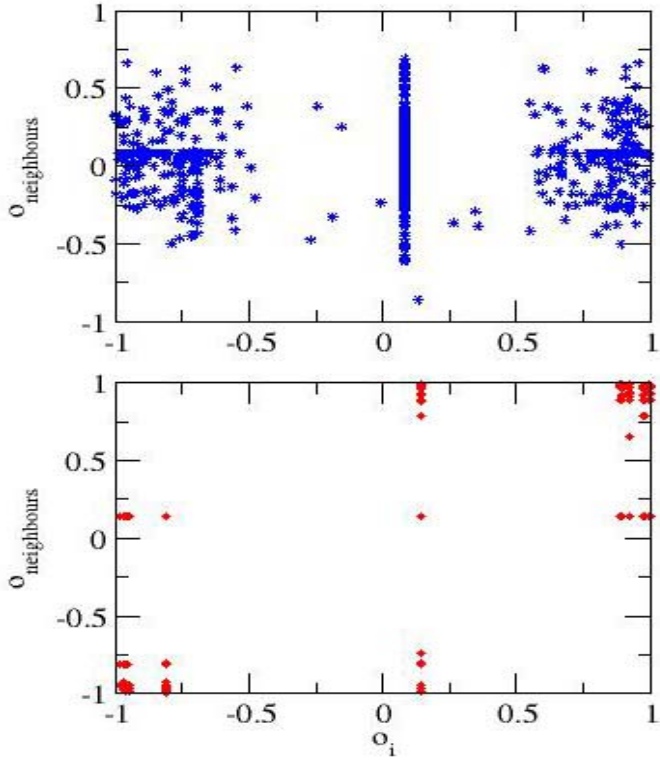

**Figure 8:** Neighbours opinion at the end of the opinion dynamics process. The three plots correspond to three different values of the parameter β. The points represent all the couples $(o_i, o)$ for all the links of the network.

### 4.1 Statistical measures

To define opinion clusters we use the procedure described in [25]: a cluster is a group of agents such that, for each couple of agents in the group, there is a chain of intermediate agents whose opinion differs less than a given threshold. We fix the threshold to be s=0.01. We measure clusters only at the end of the simulation, when a steady state is reached, and consider an opinion cluster only a group of more than one agent.

Notice that, differently by the definition of clusters used in percolation theory, in our notion of clusters no information on the topological structure is contained: we are defining opinion clusters that contains agents with very similar opinions but that a priori do not have particular neighbouring properties on the network.

The basic measure to describe the dynamical process is the *number of clusters* ($n_{cl}$). But we also need to introduce some indicators for the cluster structure. The *giant opinion cluster* is the largest cluster of agents sharing the same opinion. We will use as an indicator for the dynamics the size of the relative giant cluster size, namely, the giant cluster size normalized with the total number of agents ($g_{dim}$):

$$g_{dim} = (n_{ag} \text{ in the giant cluster}) / N_{ag}$$

If all the agents converge to the same opinion the normalized giant cluster has size $g_{dim}=1$ while, on the other extreme, if all the agents have a different opinion $g_{dim}=1/N$.

It is also useful to introduce the *average secondary cluster* dimension, that is sum of the relative average dimension of the second and third largest clusters. To estimate the indicators we will perform some experiments on a population of $N_{ag}$=1000 agents for different values of the parameter; to produce a reliable statistics, for each set of parameters, we consider 10 different initial conditions (stochastic realizations of the network) and on each of these realizations we run the opinion dynamics for 10 different times. The final result is averaged over all the 100 trials.

Figure 9 shows the result of the simulations as a function of β for fixed values of α. For all the values of α the number of clusters decreases with β: as β gets bigger (namely the extremists are more segregated at the beginning) the number of opinions available at the end of the simulation is much lower than those we had at the beginning. In particular for α < 0.9 and sufficiently large values of β, the opinions converge to uniformity. This happens because when α>0 for every agent the set of nodes with whom interaction is possible is reduced, and if an extremist is surrounded by neighbours with opinions out of the reach of its tolerance, it can not interact and reach the giant cluster. But as β is increased so is the possibility of having neighbours with similar opinions and therefore to find the way toward the giant cluster.

In Figure 9C, for α= 0.8, a characteristic behaviour of the secondary cluster size as a function of β is observed: it presents a clear maximum for β~3. The explanation is that if β is too small (β<3) extremists are set apart from each other and so is higher the possibility for them to be surrounded by neighbours with whom interaction is impossible. As a consequence, if consensus is impossible, even a large opposition cluster can not be formed.

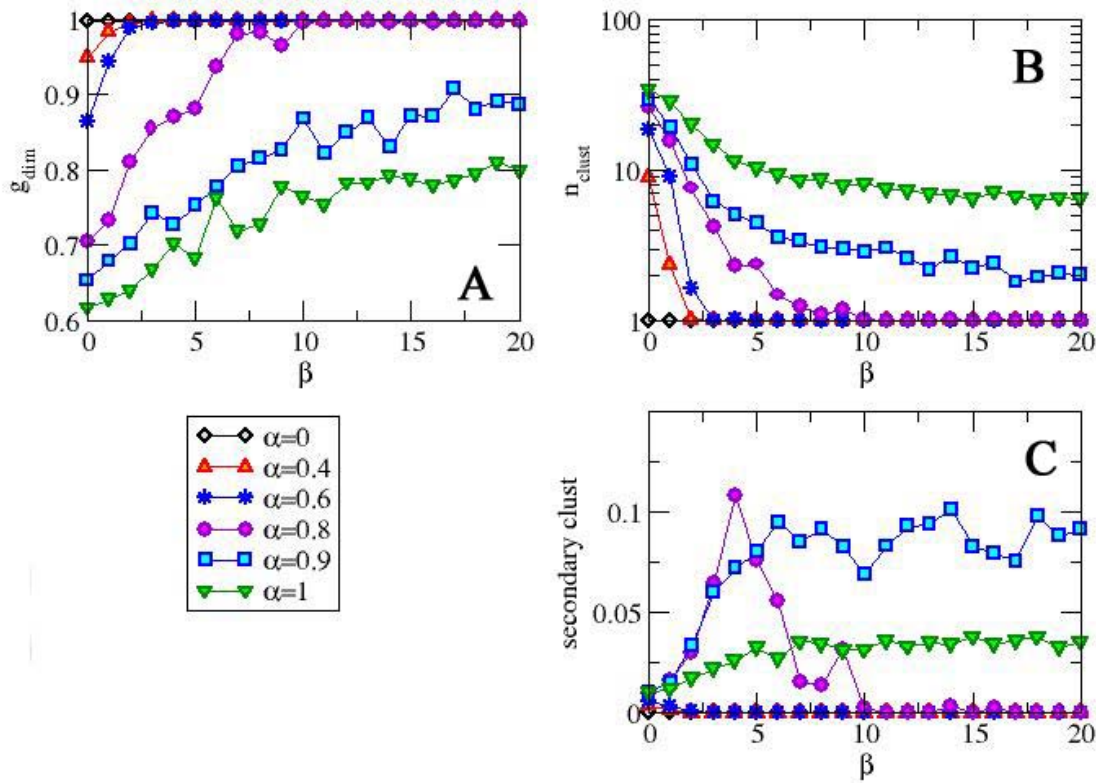

**Figure 9:** Cluster statistics for different values of α as a function of the static parameter β.

Relative size of the giant cluster (A), number of clusters (B) and average secondary cluster size (C). Results are averaged over 100 simulations.

In Figure 10, instead, we describe the behaviour of the indicators as a function of the dynamical parameter α for fixed values of β.

The number of clusters, for every value of β, increases with α, and consequently the giant cluster size decreases. For each value of β there is a critical value of the dynamical parameter α, $α_c$, such that for α<$α_c$ (when the tolerance is less depending on the opinion) the system converges to a single opinion while, for α>$α_c$ the final state shows a larger number of opinion clusters. However, we find again in Fig 10C that for low values of β the maximum size of the extremists cluster is reached for an optimal value of α. This is because when β is small (extremist randomly linked with all the other agents) and α is large (extremist tolerance is small) extremists can find again themselves surrounded by neighbours with milder opinions, with whom they can not interact, and therefore can not find the way to be connected to the cluster formed by the other extremists, while if α is low consensus is achieved.

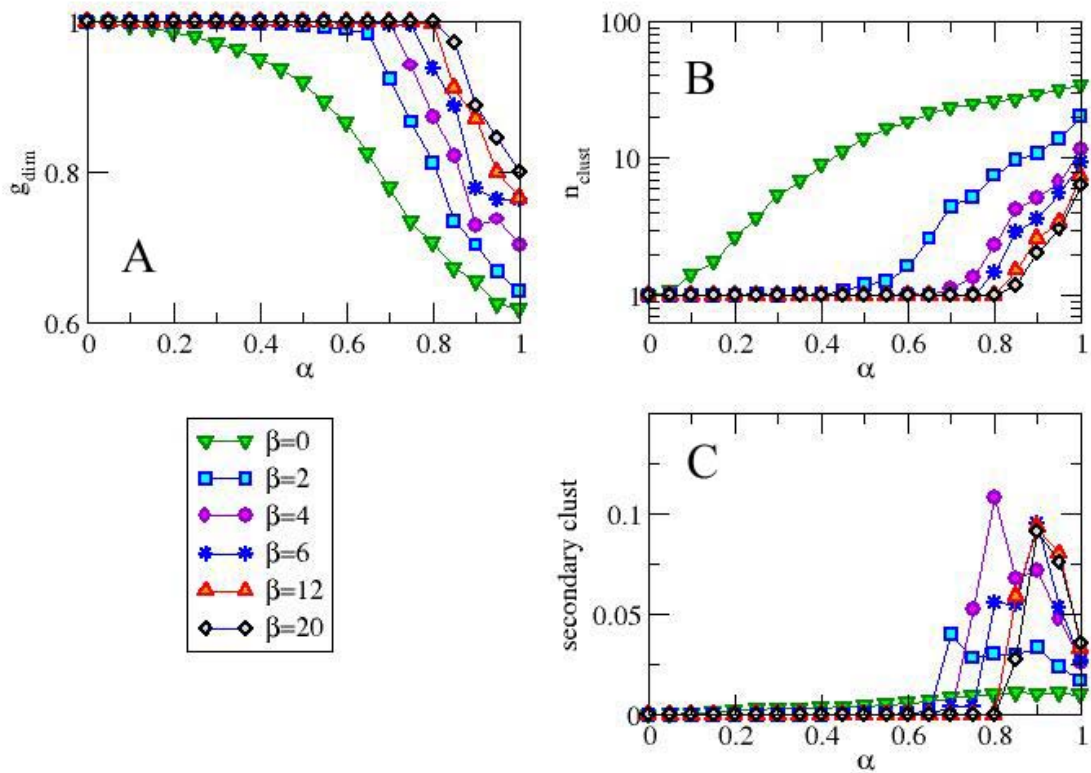

**Figure 10:** Cluster statistics for different values of β as a function of the dynamical parameter α. Giant cluster dimension (A), number of clusters (B) and extremist cluster dimension (C). Results are averaged over 100 simulations.

## 4.2 Robustness of the results changing the network topology

Some additional considerations should be added about the influence of the network topology on the result. It has been proved that in Deffuant model the presence of a scale free network does not influence the results about the phase transition from a consensus scenario to a fragmented one: independently from the network structure (random graph, lattice or scale free), and also in the case of the fully connected structure (Mean Field approach) the transition point is reached when the parameter of the bounded confidence is $\varepsilon=0.5$ [26].
Also this model presents this kind of robustness regarding to the topological choices. For these analysis We will focus on the measure of the average relative giant cluster size.
First of all we implement a mean field approach. In this case the segregation effect is not expressed, like in the case with $\beta=0$ (opinion independent scale free network). The only difference between the mean field case and the scale free network is the degree distribution. As displayed in figure 11, in the case of the mean field and in the case of a scale free network, the curves of s with respect to $\alpha$ present a very good superposition.

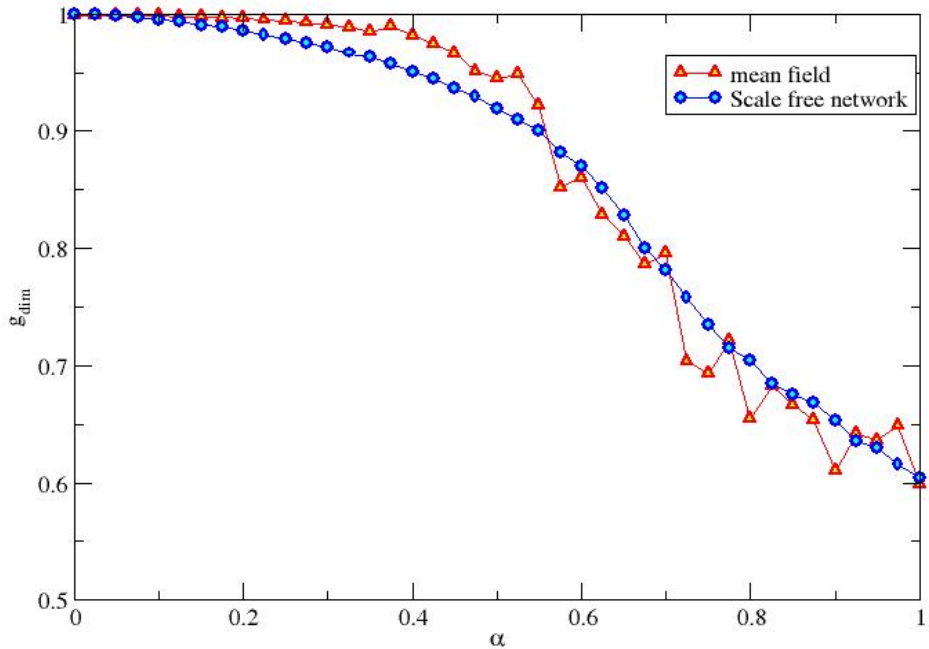

**Figure 11:** Giant cluster dimension vs $\alpha$ for a complete graph and an opinion dependent scale free structure with $\beta=0$. Each point is obtained as the average of 100 simulations with 1000 agents.

Also the presence of a scale free structure is not fundamental for the final result. In this case we perform the same analysis about the dependence of the giant cluster size from $\alpha$, for an opinion dependent network with $\beta=3$, in the scale free (1) case and in the case when the degree dependence of the connection probability is relaxed:

$$P_{(N \to i)} \approx \mathrm{Exp}[-\beta|o_N||o_N-o_i|]$$

As observed in figure 12, also in this case a very good superposition is observed.

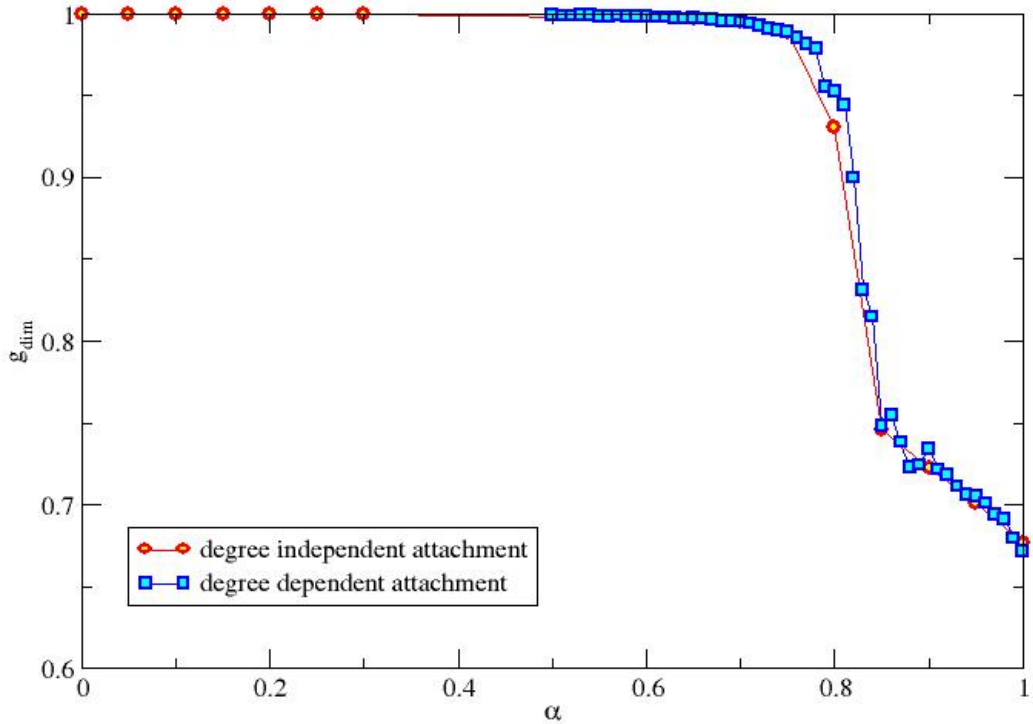

**Figure 12:** Giant cluster dimension vs α for an opinion dependent scale free structure and an opinion dependent random degree structure, both with β=3. Each point is obtained as the average of 100 simulations with 1000 agents.

## 4.3 Convergence types

In our simulations we studied the conditions needed to realize opinion uniformity. We think anyhow that in a social perspective the absence of uniformity does not automatically imply the presence of a real pluralism, i.e. the presence of many subjects able to influence the public scene. For this reason we compared in the previous section the size of the majority cluster and the size of the secondary cluster. Another possible measure of pluralism is the number of opinion clusters surviving in the stable state. According to these measure we can identify three different convergence types:

- *Uniformity*: all the agents converge to the same opinion and only one cluster is present.
- *Strong majority*: almost all the agents converge to the average opinion but few ($n_{cl} < n_{thr}$) extremist clusters remains.
- *Pluralism*: many different clusters ($n_{cl} > n_{thr}$) remain and a wide range of stable opinions is observed after the dynamical process.

We set $n_{thr} = 5$.

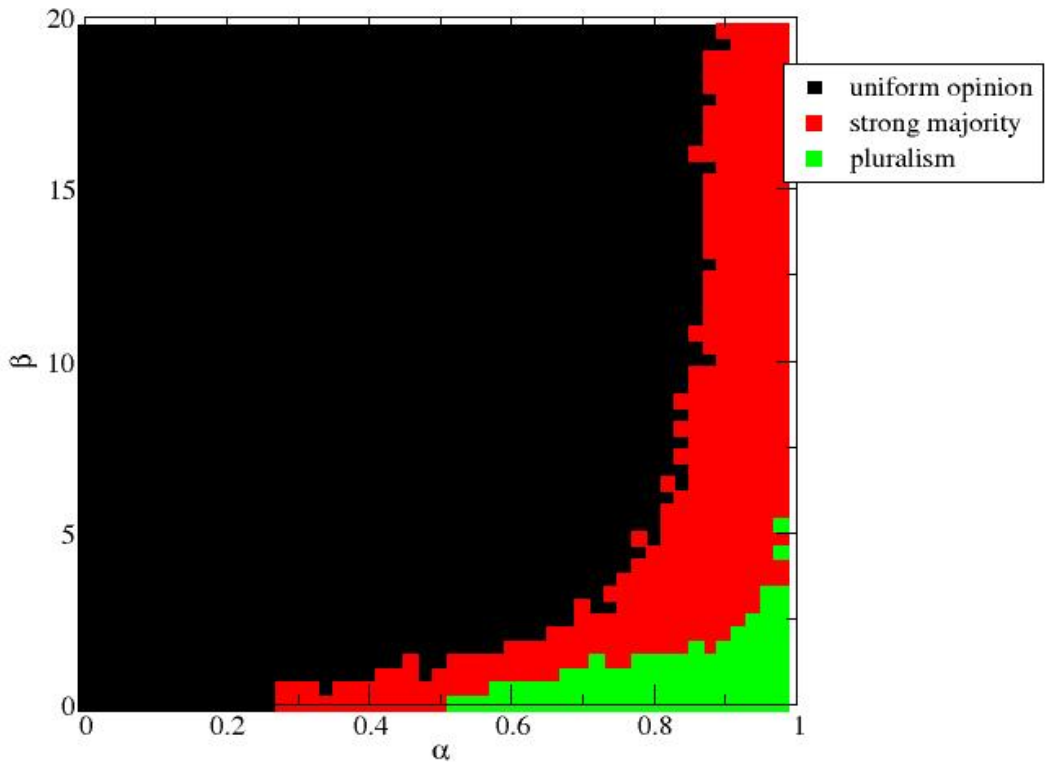

**Figure 13**: Different regimes identified by number of clusters as a function of (α,β). Each point is obtained as the average of 25 simulations with 1000 agents.

Figure 13 shows how different convergence regimes can be realized for different values of the parameter space: low values of the dynamical parameter α (α<0.2) lead to uniformity independently of the topological structure. For small values of β (β->0, namely the standard Barabasi – Albert network model), a small deviation of the dynamical model from the Deffuant's (α=0) is enough to guarantee the presence of more than one opinion at the end of the opinion evolution. On the opposite, when the network starts to exhibit a strongly segregated structure, there is no way to avoid the uniformity apart from very high values of α (α>0.8). When a treshold value of β (~10), is crossed, the transition between the uniform and the strong majority regimes is realized at the same value of α, the critical value $α_C$ that we identified in Figure 10.

A situation where a large number of opinion remains available (pluralism) can be realized only in a very particular case: small β and large α. This convergence type can be observed in Figure 10C: for β=2, this is the α-range where the secondary cluster size starts to decrease after a flat zone corresponding to the strong majority area.
In these conditions the extremist agents are neither able to interact with all their neighbours that, since β is small can, have very different opinions, neither to create minority clusters. This situation is the optimal one for the opinion preservation since it leads to the formation of clusters different from the majority consensus even for non-extreme opinions.

# 5-Conclusions

Different social dynamics arise from the implementation of our algorithms of segregation.
A non trivial result is that opinion dependent homophily, i.e. the fact that extremists prefer to form links with people sharing similar ideas while neutrals do not consider the opinion of other people while forming a link, is in fact not preventing the formation of a uniform consensus, but only making its realization slower (Fig.4). On the contrary, when also opinion dependent tolerance and sensitivity are introduced, the topological segregation actually helps the formation of a large majority cluster (Fig.5). Opinion dependent homophily and opinion dependent sensitivity are two mechanisms that seem to work in the same direction, i.e. forming isolated clusters of extremists, but when combined neutralize each other and result in the reduction or even in the absence of extremists clusters. For instance, in Figure 13, if we set $\beta=0$ and we increase $\alpha$ up to 0.6 the system goes from uniformity to strong majority to pluralism, but if at this point we increase $\beta$, increasing the starting segregation of the opinions, we come back first to strong majority and then to uniformity. This counter-intuitive result is easy to understand if one thinks that the tendency to form links with people similar to you ensure that you are never going to be isolated, and in this way the dialogue can continue to shape the different opinions and to make them closer.

Making the extremists less tolerant and sensitive to different opinions predictably results in a reduced size of the majority cluster. If the process is pushed too far extremists do not form even minority cluster and remain simply isolated. The formation of self-referring opinion fringes is indeed a very common phenomenon. The price these small groups pay for their isolation is simply the impossibility to efficiently influence the society. In this perspective the most interesting feature is the one we see in Fig.10C: for low values of $\beta$ there is an optimal value of $\alpha$ maximizing the size of minority clusters. We can see this combination of parameters as the rise of a "second opinion" (and maybe third, fourth, and so on) effectively separated from the majority consensus and at the same time able to affect society thanks to its size.

Due to the initial opinion distribution and to the fact that agents with a more neutral opinion have a larger range of interaction, nodes on average will weaken their opinions at each step for any value of the parameters. What prevents then the formation of a uniform consensus for some parameter values? If at any moment there is a node whose neighbours with a milder opinion are all at a distance larger than its tolerance, this node will be forever isolated from the majority, and uniform consensus will never be reached. We saw that increasing $\beta$ the possibility for this to happen is decreasing, because neighbours will in most cases share similar opinions, while on the opposite increasing $\alpha$ will decrease the tolerance ranges and make isolation more probable. For this reason, for sufficiently high $\alpha$ there is a critical value of $\beta$ for which isolation appears and uniformity breaks down, and this value is rapidly diverging when $\alpha$ is increased. This phase transition, happening when the opinion space is not "fully connected" anymore, happens at the border of the black and red areas in Fig.13. Furthermore, we can see also the formation of minority clusters in the same way, since it depends on whether the set of the extremists is fully connected or not.

The influence of minorities in the construction of public opinion has been thorough analyzed by sociologist, using empirical methods. In 1969 an experiment was performed about how the response in the perception of a color of a majority changes in presence of a minority [27]. This and other subsequent experiments confirmed that the efficiency of minority influence definitely depends on the behavioural styles of the minorities.
In particular Moscovici theory of minority influence [28,29] is based on the fact that a minority source (that never yield or compromise) enhances influence in a debate if it demonstrates to be always consistent: its consistency conveys information to its position and it is able to generate in the recipients a higher level of conflict with the majority source.

This is comparable with our result that a non-consensus situation, where minority clusters are present, can be realized only after the critical point in $\alpha$, the parameter that tunes the level of inflexibility of the extremists.

An other important aspect about the negative back-reaction of a too strong consistency is underlined in [29-31]: when the recipients perceive in the minority a "dogmatic" attitude, they will be less disposed to debate. This assumption is again in good agreement with our model where a optimal value of $\alpha$ was identified for the creation of cohesive minorities clusters.

Apart from consistency, an other factor results to be important for the influence of minority sources: their dissimilarity from the majority target. A clear distinction is present between the influence of in-group and out-group minorities [32,33]. In this sense, the so called "deviant" minorities, those which are not considered part of the society, face much higher difficulties to gain consensus. This concept is represented in our model by the fact that the non-consensus phase is much easier to be reached for low values of the segregation parameter $\beta$.

Finally, in this paper we hypothesized that extremists with distant opinions are not interacting. What would be interesting is to implement a negative interaction, i.e. an interaction between extremists that is reinforcing, instead of weakening, the respective opinions. A very common example is the reaction mechanism with whom an opinion minority tolerated by the common consensus becomes more segregated if tolerance turns into widespread hostility.

Another assumption that we did in this paper is that the opinion dynamics happens rapidly so that we can consider the social network as a static framework. On the other side, to consider a slow process where the agents adapt their opinions according to the social background, we should keep into account the feedback mechanism that the dynamical process has on the network structure itself [34,35]. In forthcoming projects we will consider what changes in the opinion dynamics when the social relationships naturally evolve in time according to the opinion changes of the agents.

**Acknowledgments:**

The authors would like to thank A. Barrat, S.Fortunato and J.J. Ramasco for helpful discussions and for the very important suggestions.

**References:**

1. Golding P, Harris P (1997) Beyond Cultural Imperialism: Globalization. Communication and the New International Order, Londres: Sage.

2. Jenkins H, Thorburn D (2003) Democracy and New Media.

3. Castellano C, Fortunato S, Loreto V (2007) Statistical physics of social dynamics. eprint arXiv: 07103256v1.

4. Clifford P, Sudbury A (2003) A model for spatial conflict. Biometrika 60: 581-588.

5. Holley RA, Liggett TM (1975) Ergodic Theorems for Weakly Interacting Infinite Systems and the Voter Model. The Annals of Probability 3: 643-663.

6. Axelrod R (1997) The Dissemination of Culture: A Model with Local Convergence and Global Polarization. The Journal of Conflict Resolution 41: 203-226.

7. Deffuant G, Neau D, Amblard F, Weisbuch G (2000) Mixing beliefs among interacting agents. Advances in Complex Systems 3: 87-98.

8. Weisbuch G, Deffuant G, Amblard F, Nadal JP (2002) Meet, discuss, and segregate! Complexity 7: 55-63.

9. Deffuant G, Amblard F, Weisbuch G, Faure T (2002) How can extremism prevail? A study based on the relative agreement interaction model. Journal of Artificial Societies and Social Simulation 5: 1.

10. Galam S, Jacobs F (2007) The role of inflexible minorities in the breaking of democratic opinion dynamics. Physica A: Statistical Mechanics and its Applications 381: 366-376.

11. Stauffer D, Meyer-Ortmanns H (2004) Simulation of Consensus Model of Deffuant et al. on a BARABÁSI-ALBERT Network. International Journal of Modern Physics C 15: 241-246.

12. Schelling TC (1969) Models of Segregation. The American Economic Review 59: 488-493.

13. Stauffer D, Solomon S (2007) Ising, Schelling and self-organising segregation. The European Physical Journal B-Condensed Matter and Complex Systems 57: 473-479.

14. Centola D, Gonzalez-Avella JC, Eguiluz VM, San Miguel M (2007) Homophily, Cultural Drift, and the Co-Evolution of Cultural Groups. Journal of Conflict Resolution 51: 905.

15. Kandel DB (1978) Homophily, Selection, and Socialization in Adolescent Friendships. The American Journal of Sociology 84: 427-436.

16. McPherson M, Smith-Lovin L, Cook JM (2001) Birds of a Feather: Homophily in Social Networks. Annual Review of Sociology 27: 415-444.

17. Galam S (2005) Heterogeneous beliefs, segregation, and extremism in the making of public opinions. Physical Review E 71: 46123.

18. Barabási AL, Albert R (1999) Emergence of Scaling in Random Networks. Science 286: 509.

19. Bianconi G, Barabasi AL (2001) Competition and multiscaling in evolving networks. Europhysics Letters 54: 436-442.

20. Shelling T (1971) Dynamic models of segregation. Journal of Mathematical Sociology 1: 143-186.

21. Fortunato S, Barthelemy M (2007) Resolution limit in community detection. Proceedings of the National Academy of Sciences 104: 36.

22. Newman MEJ, Girvan M (2004) Finding and evaluating community structure in networks. Physical Review E 69: 26113.

23. Ashforth BE, Mael F (1989) Social identity theory and the organization. Academy of Management Review 14: 20-39.

24. Turner JC, Brown RJ, Tajfel H (1979) Social comparison and group interest in ingroup favouritism. European Journal of Social Psychology 9: 187-204.

25. Deffuant G (2006) Comparing Extremism Propagation Patterns in Continuous Opinion Models. Journal of Artificial Societies and Social Simulation 9: 8.

26. Fortunato S (2004) Universality of the Threshold for Complete Consensus for the Opinion Dynamics of Deffuant et al. International Journal of Modern Physics C 15: 1301-1307.

27. Moscovici S, Lage E, Naffrechoux M (1969) Influence of a consistent minority on the responses of a majority in a color perception task. Sociometry 32: 365-380.

28. Moscovici S (1980) Toward a theory of conversion behavior. Advances in experimental social psychology 13: 209-239.

29. Moscovici S (1985) Social influence and conformity. Handbook of social psychology 2: 347–412.

30. Maass A, Clark RO (1984) Hidden impact of minorities. Psychological Bulletin 95: 428–450.

31. Nemeth C, Wachtler J (1973) Consistency and modification of judgment. Journal of Experimental Social Psychology 9: 65-79.

32. Clark RD, Maass A (1988) The role of social categorization and perceived source credibility in minority influence. European Journal of Social Psychology 18: 381-394.

33. Clark RD, Maass A (1988) Social categorization in minority influence: The case of homosexuality. European Journal of Social Psychology 18: 347-364.

34. Kozma B, Barrat A (2007) Adaptive networks: the example of consensus formation. American Physical Society, APS March Meeting, March 5-9, 2007, abstract# 22: 010.

35. Kozma B, Barrat A (2008) Consensus formation on adaptive networks. Physical Review E 77: 16102.